Seventeen 2 Micron All Sky Survey (2MASS) hypervelocity stars (HVS) from Gaia DR3

Mudumba Parthasarathy

Indian Institute of Astrophysics, Koramangala 2nd block, Bangalore 560034, India

Abstract

As part of an ongoing search for hypervelocity stars (HVS) I found seventeen two micron all sky survey (2MASS) sources with Gaia G magnitudes less than 16.0 and radial velocities less than -600 km/sec. All these stars are brighter in the K band when compared with their V and G magnitudes. Ten of these (including three carbon stars) are long period variable stars (LPV) of Mira type. One is a relatively nearby high proper motion star and one is a very high galactic latitude chemically peculiar metal-poor star. It may be a galactic halo star. One star is a Kepler red giant, two stars may be cluster members and two are in the star forming region (probably YSOs). It is not clear how these stars acquired such high radial velocities. Further study of these seventeen stars is needed.

1. Introduction

I have searched the Gaia DR3 data (Gaia Collaboration 2022a, b) for HVS adopting a simple search criteria. This paper is a continuation of that work. For more details of search criteria see Parthasarathy (2023, 2024) and there is no need to repeat them here. In this research note I present seventeen 2MASS sources with Gaia DR3 G magnitudes less than 16.0 and radial velocities less than -600 km/sec that I found (Table 1). With this magnitude and velocity criteria mentioned in the abstract I have found only 17 2MASS stars. Other types of HVS found with the same search criteria were presented in Parthasarathy (2023).

Table 1.  Selected 2MASS HVS

| 2MASS | l | b | mu | G | RV | d | U | V | W | Vt |
|---|---|---|---|---|---|---|---|---|---|---|
| | deg | deg | mas/yr | mag | km/sec | pc | ----- | km/sec | --------------- | |
| J17500025-1542221 | 11.9 | 5.9 | 9.83 | 14.08 | -633 | 4833 | -617 | -130 | -66 | 225 |
| J18213211-1740598 | 13.8 | -1.6 | 8.99 | 15.98 | -712 | 7836 | -691 | -170 | 20 | 334 |
| J15003649-5529145 | 320.6 | 2.9 | 6.68 | 15.74 | -754 | 7335 | -581 | 478 | -38 | 232 |
| J17322853-3342336 | 354.5 | -0.2 | 6.32 | 15.08 | -632 | 5581 | -629 | 61 | 2 | 167 |
| J17502866-3242362 | 357.3 | -2.8 | 2.31 | 15.73 | -740 | 7770 | -738 | 35 | 37 | 85 |
| J17563584-2152290 | 7.3 | 1.5 | 9.57 | 15.82 | -620 | 9184 | -615 | -79 | -16 | 416 |
| J18084725-1403252 | 15.6 | 2.8 | 7.84 | 15.74 | -805 | 8127 | -775 | -216 | -39 | 302 |
| J18271007-1907365 | 13.2 | -3.5 | 8.53 | 14.68 | -682 | 6582 | -662 | -155 | 41 | 266 |
| J18443216+0253312 | 34.7 | 2.9 | 6.54 | 15.88 | -792 | 7604 | -650 | -450 | -39 | 236 |
| J19283045+1312303 | 48.9 | -2.0 | 6.02 | 15.19 | -804 | 5516 | -528 | -605 | 28 | 157 |
| J07383691+3212208 | 187.4 | 23.4 | 112.05 | 13.77 | -689 | 187 | 627 | 82 | -273 | 100 |
| J19352291+4626156 | 79.1 | 12.3 | 3.78 | 14.26 | -620 | 6169 | -114 | -594 | -132 | 111 |
| J11220124+5309490 | 149.3 | 59.1 | 4.78 | 13.20 | -800 | 10722 | 353 | -209 | -686 | 243 |
| J23470671+6311571 | 115.8 | 1.2 | 3.37 | 15.14 | -828 | 3150 | 360 | -745 | -18 | 50 |
| J00504461+5824384 | 122.8 | -4.5 | 2.12 | 14.63 | -723 | 4152 | 391 | -606 | 56 | 42 |
| J02274456+6140299 | 134.1 | 0.9 | 0.65 | 15.31 | -794 | 2514 | 553 | -570 | -13 | 8 |
| J02475329+6044285 | 136.7 | 1.0 | 2.32 | 11.63 | -657 | 1370 | 478 | -451 | -12 | 15 |

2. Discussion

From the data given in Table 1 I calculated the U, V, W, and Vt (tangential velocity) velocities and they are given in Table 1. These velocities and radial velocities (Table 1) clearly indicate that these seventeen 2MASS stars are HVS. Notes on these seventeen HVS 2MASS sources are given below. For 10 LPVs the radial velocities given in Table 1 are from Gaia DR3 ( Gaia Collaboration 2022a,b). The typical errors in Gaia DR3 velocities of these 10 LPVs are of the order of 10 km/sec. For five stars the radial velocities are from the catalogue of Jonsson et al. (2020) which were derived from the high resolution spectra and the errors are of the order of 1.0 km/sec. The distances given in Table 1 are from Bailer-Jones et al. (2021). For the first ten stars the light curves are given in the Gaia DR3 (The second catalogue of LPV candidates (Lebzelter et al. 2023)). The Teff, log g, and [Fe/H] values given below for some of the stars are from the Gaia DR3 data (Gaia Collaboration 2022a, b).

2MASS J17500025-1542221 (OGLE BLG-LPV-251783 = Gaia DR3 4148047120720805760). It is a M-type long period Mira variable star with a period of 244.991098 days, V maximum = 11.014 and V minimum = 12.921 (Lebzelter et al. 2023 and Iwanek et al. 2022). Gaia DR3 light curve and spectrum are available.

2MASS J18213211-1740598 (OGLE BLG-LPV-262377 = Gaia DR3 4096520535498750976). It is a 254.915131 days long period Mira variable (Lebzelter et al. 2023 and Iwanek et al. 2022). V maximum = 13.588 and V minimum = 15.401. Gaia DR3 light curve is available.

2MASS J15003649-5529145 (Gaia DR3 5881574502067686016). Gaia DR3 light curve gives a period of 142.198625 days. Gaia DR3 data gives Teff = 4384K, log g = 1.523 and [Fe/H] = -0.724.

2MASS J17322853-3342336 (Gaia DR3 4054394740276777856). It as a carbon star and a long period variable candidate. Gaia DR3 light curve is available. It may be a carbon Mira.

2MASS J17502866-3242362 (Gaia DR3 4043485523412935040). Gaia DR3 data classify it as a long period variable candidate and as a carbon star. Gaia DR3 light curve is available. It may be a carbon Mira.

2NASS J17563584-2152290 (Gaia DR3 4070380230599079552). It is a long period variable candidate. Gaia DR3 data gives Teff = 3663K, log g = 1.026 and [Fe/H] = -0.440.

2MASS J18084725-1403252 (Gaia DR3 4146964342221788160). It is classified as a long period variable candidate and as a carbon star. It may be a carbon Mira. Gaia DR3 data gives Teff = 3523K, log g = 0.557, and [Fe/H] = -0.247.

2MASS J18271007-1907365 (Gaia DR3 4093092155089523200). It is classified as a long period variable candidate. Gaia DR3 data gives Teff = 3407K, log g = 0.244, and [Fe/H] = -0.62.

2MASS J18443216+0253312 (Gaia DR3 4280188249119889536). It is classified as a long period variable candidate. Gaia DR3 data gives Teff = 3882K, log g = 0.178, and [Fe/H] = -1.1

2MASS J19283045+1312303 (Gaia DR3 4316462477768150144). It is a long period (240.137737 days) variable candidate. Gaia DR3 data gives Teff = 3265K, log g = 0.361, and [Fe/H] = 0.35

2MASS J07383691+3212208 (Gaia DR3 892417483109166976). It is a high galactic latitude and high proper motion star (Table 1). It is a variable star with a period of 2.259466 days (Hartman et al. 2011). it may be a  rotating or ellipsoidal variable. It is relatively a nearby star (distance 187 pc (Table 1)). Gaia DR3 data gives Teff = 3783K, log g = 4.535, and [Fe/H] = -1.004. It appears to be a metal-poor red dwarf star. Gaia DR3 spectrum is available. High resolution spectroscopy of this star is important.

2MASS J19352291+462656 (Gaia DR3 2128166023276246400). Its radial velocity given in Table 1 is from Frasca et al. (2016). It is a rapidly rotating (vsini = 125 km/sec) and high galactic latitude G5 III Kepler red giant. Frasca et al. (2016) derived Teff = 5306K, log g = 3.13 and [Fe/H] = -1.08. Yu et al. (2018) derived Teff = 4969K, log g = 2.339, and [Fe/H] = 0.07. Gaia spectrum is available. It may be a close binary. [Fe/H] needs to be confirmed.

2MASS J11220124+5309490 (Gaia DR3 839759740951501952). It is a very high galactic latitude (Table 1) chemically peculiar star. In SIMBAD its Teff = 4241K, log g = 1.152, and [Fe/H] = -1.114. Ting et al. (2019) derived Teff = 4413K, log g = 2.229, and [Fe/H] = -1.224. It is most likely a galactic halo metal poor hypervelocity red-HB or post-AGB or a red giant.

2MASS J23470671+6311571 (Gaia DR3 2016055454027115520). It may be a member of galactic cluster (Spina et al. 2021). Gaia DR3 light curve shows that it is a variable star and Teff = 5879K, log g = 3.662.

2MASS J00504461+5824384 (Gaia DR3 424426817070254208). It is a red giant in an open cluster ( Warren and Cole 2009). Gaia DR3 data gives Teff = 4927K, log g = 2.796, and [Fe/H] = -0.348.

2MASS J02274456+6140299 (Gaia DR3 513579686610504064). It is a B3-6 Ve star in star forming complex W3. Its spectrum shows strong H-alpha emission line (Kiminki et al. 2015, Jose et al. 2016, Fratta et al. 2021). Gaia DR3 data gives Teff = 6665K, log g = 2.61, and [Fe/H] = -3.11. This Teff value is not in agreement with the spectral type. The V-K colour indicates IR excess. In SIMBAD it is classified as a YSO. It is not clear how such a young massive star in star forming region acquired such high radial velocity. Redetermination of radial velocity and confirmation of the membership of star forming complex W3 is need to further understand this puzzle.

2MASS J02475329+6044285 (Gaia DR3 464830536494375808). It is a B6e star in star forming region W5 (Koenig and Allen 2011). It may be an open cluster member. The V-K colour indicates IR excess. What is stated above in the case of 2MASS J02274456+6140299 holds good for this YSO 2MASS J02475329+6044285. These two stars needs further study to establish whether or not they are HVS and if they are true HVS then how they acquired such high radial velocities is a puzzle. There is not much literature and data on these two stars to resolve the puzzle.

The stellar winds of the early and late type HVS may form tail like feature opposite to the direction of their motion and the mass-loss rate of HVS may be little bit more than that of normal stars with low radial velocities and same spectral types.

3. Acknowledgements

This research has made use of the Gaia DR3 and SIMBAD databases, operated at CDS, Strasbourg, France and I have used the NASA ADS. No funding for this research. I am thankful to Prof. Chris Lintott for helpful comments.